\documentclass[prb,superscriptaddress,floatfix,twocolumn,showpacs,amsmath,amssymb,final]{revtex4}
\usepackage{graphicx} 
\usepackage{dcolumn} 
\usepackage{bm} 

\begin{document}

\title{Spin-flop transitions and spin-wave gaps in La$_2$CuO$_4$}

\author{Andreas L\"uscher}
\affiliation{School of Physics, University of New South Wales, Sydney 2052, Australia}
\author{Oleg P. Sushkov}
\affiliation{School of Physics, University of New South Wales, Sydney 2052, Australia}

\date{\today}

\begin{abstract}
We study the spin-wave spectrum and the spin-flop transitions in La$_2$CuO$_4$ in a uniform magnetic field at zero temperature. Using the non-linear $\sigma$-model, we show that a field applied along the orthorhombic $b$ direction leads to a two-step rotation of the staggered magnetization, first in the $bc$ and then in the $ac$ plane, until the order parameter is completely aligned along the $c$ axis. In contrast, for a perpendicular magnetic field, we find a conventional spin-flop transition induced by the competition between the field and the interlayer coupling. A comparison with recent measurements of the field-dependence of the in-plane spin-wave gap shows a beautiful agreement between theory and experiments. 
\end{abstract}


\maketitle

\section{Introduction}
Since the discovery of high-temperature superconductivity~\cite{bednorz86} in materials derived from the La$_2$CuO$_4$ (LCO) parent compound, much progress has been made towards a better understanding of the magnetic properties of this Mott insulator. From a theoretical point of view, the puzzle behind the physics of the cuprates lies in the interplay between the charge carriers and the spin background. A sound knowledge of the magnetic properties of the parent compound is therefore indispensable for a successful description of the doped materials. The relatively simple structure of LCO facilitates this task considerably by allowing one to perform a great number of experiments.

At low-energy, the charge degrees of freedom in LCO are frozen and the remaining copper spins are well described by an antiferromagnetic spin-$1/2$ Heisenberg model. The field-theoretical treatment of this model with isotropic superexchange interactions was introduced by Chakravarty {\it et al.}\cite{chakravarty88,chakravarty89,chubukov94} and proved to be very successful in describing the paramagnetic phase of LCO. However, it was noted early on that in the low-temperature orthorhombic phase, anisotropies due to the Dzyaloshinskii-Moriya (DM) and the direct-exchange interaction cannot be neglected~\cite{shekhtman92,koshibae94} and greatly influence the magnetic behavior in this phase. Quite recently, Chovan and Papanicolaou~\cite{chovan00} as well as Silva Neto {\it et al.}~\cite{silvaneto06} generalized the non-linear $\sigma$-model (NLSM) approach to systems with anisotropic superexchange. Employing this formalism, we present a study of the evolution of the magnetic properties of LCO in the presence of a uniform magnetic field, which allows us to compare our findings with recent measurements of the spin-wave gaps determined by Raman spectroscopy~\cite{gozar04} and shed some light on the controversy concerning the nature of the spin-flop transitions. Although the mechanism of the spin-flop transition observed for a perpendicular (to the copper-oxide planes) magnetic field is well understood~\cite{thio88}, and basically due to the competition between the antiferromagnetic interlayer coupling and the magnetic field favoring ferromagnetic alignment, the precise value of the critical field is not yet univocally determined~\cite{gozar04,reehuis06}. We hope that the present work encourages new neutron scattering experiments that can clarify this issue. Even more controversial is the case of a uniform field applied along the easy axis, i.e., along the direction of the staggered magnetization observed at zero field. There is no doubt that in this case the order parameter is rotated towards the $c$ axis, but Thio {\it et al.}~\cite{thio90} suggested a two-step reorientation (supported by our calculations) while Ono and co-workers~\cite{ono04} argued in favor of a single continuos rotation. 

The rest of the paper is organized as follows. In Sec.~\ref{sec:model}, we introduce the non-linear $\sigma$-model describing LCO. We first examine this model in the absence of a magnetic field and discuss the strength of the interlayer coupling and the renormalization of the gyromagnetic ratio due to the spin-orbit coupling. Sec.~\ref{sec:spectrum} then contains our main results of the evolution of the spin-wave spectrum in the presence of uniform magnetic fields applied along the three different crystal axes. These results allow us to proceed to a careful comparison between our findings and recent experimental observations of one-magnon Raman spectra~\cite{gozar04} in Sec.~\ref{sec:gaps}. Finally, we present our conclusions in Sec.~\ref{sec:conclusion}.

\section{Model \label{sec:model}}
The NLSM is a very convenient tool to describe the low-energy dynamics of the Heisenberg model, relevant for LCO. In this approach, the staggered component of the copper spins in a \emph{single layer}  of LCO is represented by a continuous vector field ${\vec n}\left({\bf r}\right)$ of unit length ${\vec n}\left({\bf r}\right)^2=1$. To avoid confusion, we denote vectors acting in the three-dimensional (3D) spin space by arrows and vectors acting in the 2D coordinate space by the usual bold font. We consider a system of $L$ layers and indicate a given layer by a subindex $l$, i.e. ${\vec n}_{l}\left({\bf r}\right)$.  Throughout this paper, we adopt the orthorhombic coordinate system, with unit vectors ${\vec e}_{\alpha}$, $\alpha=a,b,c$. Coordinate and spin space are linked through the pinning of the  commensurate N\'eel magnetization to the orthorhombic $b$ axis. It is therefore convenient to use the same coordinate system in both cases with real space unit vectors ${\bf e}_{\alpha}={\vec e}_{\alpha}$.
According to Refs.~\onlinecite{chovan00,silvaneto06},  the Lagrangian reads
\begin{multline} \label{eq:lagrangian}
{\cal L} = \frac{\chi_\perp}{2} \sum_{l=1}^L \int d^2r \left\{ \dot{\vec n}_{l}^2-2 \dot{\vec n}_{l} \left( {\vec n}_{l} \times {\vec B}\right) + \left( {\vec n}_{l}\times{\vec B} \right)^2 \right. \\ \left. -c^2 \left({\bf \nabla}{\vec n}_{l}\right)^2 - D^2 \left(n_{l}^a \right)^2 - \Gamma_c  \left(n_{l}^c\right)^2  \right. \\ - \left. 2{\vec B} \cdot \left({\vec D}\times {\vec n}_{l}\right)- \gamma^2 {\vec n}_{l} \cdot {\vec n}_{l+1} \right\}
\ ,
\end{multline}
with the ``hard'' constraint~\cite{comment1} ${\vec n}\left({\bf r}\right)^2=1$. Here $\chi_{\perp}\approx\frac{1}{16 J}\approx 4.5 \cdot 10^{-4} \left(meV\right)^{-1}$ is the magnetic susceptibility of the Heisenberg model and $c=1.17 \sqrt{2} J\approx 230 \ meV$ the spin-wave velocity (using the exchange coupling $J\approx 140 \ meV$ given in Ref.~\onlinecite{keimer92}). Note that due to quantum fluctuations in the Heisenberg model, $\chi_{\perp}$ is reduced by $50 \%$ compared to its classical value. The anisotropies are due to the DM interaction, with a DM vector directed along the orthorhombic $a$ direction of length $D \approx 2.5 \ meV$, and the XY-term, which leads to $\sqrt{\Gamma_c} \approx 5 \ meV$. These values follow from neutron scattering~\cite{keimer93}. The magnetic field ${\vec B}$ is given in units of $g \mu_{B}=g \ 5.79 \cdot 10^{-2} \ meV/T$, $g$ being the gyromagnetic ratio and $\mu_{B}$ the Bohr magneton. The coupling between spins in adjacent copper-oxide layers is described by the last term in Eq.~(\ref{eq:lagrangian}). The coupling constant $\gamma^2$ can be determined via the spin-wave dispersion (see Sec.~\ref{sec:coupling}) or the critical field of the spin-flop transition (see Sec.~\ref{sec:fieldc}).

\subsection{Interlayer coupling constant\label{sec:coupling}}
Let us first study the Lagrangian~(\ref{eq:lagrangian}) in the absence of a magnetic field and determine the spin-wave dispersion in order to obtain an estimate of the strength of the interlayer coupling constant $\gamma^2$. In the static limit, the staggered magnetization is oriented along the orthorhombic $b$ direction (due to the anisotropies) and we can therefore consider excitations of the form
\begin{equation*}
{\vec n}_{l}\left({\bf r}\right) =
\begin{pmatrix} 
n_{l,1}\left({\bf r}\right) \\[0.2cm]
\left(-1\right)^l \sqrt{1-n_{l,1}\left({\bf r}\right)^2-n_{l,2}\left({\bf r}\right)^2} \\[0.2cm]
n_{l,2}\left({\bf r}\right)
\end{pmatrix} \ .
\end{equation*}
For small deviations from this ground state, the time derivative is given by
\begin{equation*}
\dot{\vec n}_{l}\left({\bf r}\right) =
\begin{pmatrix} 
\dot{n}_{l,1}\left({\bf r}\right) \\
0 \\
\dot{n}_{l,2}\left({\bf r}\right)
\end{pmatrix} \ .
\end{equation*}
Introducing the Fourier transformation
\begin{equation} \label{eq:fourier}
n_{l,j} \left({\bf r}\right) = \frac{1}{\sqrt{L}} \sum_{q} \int \frac{d^2 k}{\left(2\pi\right)^2} \ e^{i \left(q l+ {\bf k}Ê\cdot {\bf r}\right)} n_{{\bf k},q,j}
\end{equation}
and substituting these  expressions in the Lagrangian~(\ref{eq:lagrangian}), we find
\begin{multline*}
{\cal L}=\sum_{j=1,2}\sum_{q} \int \frac{d^2 k}{\left(2\pi\right)^2} \ \left\{ \dot{n}_{{\bf k},q,j}^* \dot{n}_{{\bf k},q,j} \right. \\ \left. - n_{{\bf k},q,j}^* n_{{\bf k},q,j} \left[c ^2 k^2+ M_{j}^2 + \gamma^2\left(1+\cos q\right)\right] \right\} \ ,
\end{multline*}
with $M_{1}^2=D^2$ and $M_{2}^2=\Gamma_{c}$. Since the prefactor $\chi_{\perp}/2$ and any terms independent of ${\vec n}_{l}\left({\bf r}\right)$ are not important for the purpose of the present work, we omit them from now on. The equations of motion read
\begin{equation*}
\ddot{n}_{{\bf k},q,j} =  -\left[c ^2 k^2+ M_{j}^2 + \gamma^2\left(1+\cos q\right)\right] n_{{\bf k},q,j} \ ,
\end{equation*}
resulting in the following spin-wave spectrum
\begin{align} \label{eq:spectrum0}
\omega_{1}^2 &= c^2 k^2+D^2+ \gamma^2\left(1+\cos q\right) \ , \nonumber \\
\omega_{2}^2 &= c^2 k^2+\Gamma_{c}+ \gamma^2\left(1+\cos q\right) \ ,
\end{align}
with the usual time dependence 
\begin{equation} \label{eq:timedep}
n_{{\bf k},q,j}\left(t\right) = e^{i \omega t} n_{{\bf k},q,j}^0 \ .
\end{equation}
Both, the in- and out-of-plane modes are gapped, due to the presence of the anisotropies $M_{j}$. Comparing Eqs.~(\ref{eq:spectrum0}) with the spin-wave dispersion given in Eq.~(2) of Ref.~\onlinecite{keimer92}, we find
\begin{equation} \label{eq:gamma}
\gamma^2=2 \alpha_{\perp} c^2\approx 5 \ \left(meV\right)^2 \ ,
\end{equation}
using $\alpha_{\perp}=5 \cdot 10^{-5}$. Although a comparison with the experimentally determined spin-wave spectrum is theoretically possible, we would like to point out that the above value of $\alpha_{\perp}$ is actually based on estimates of the spin-flop transition presented in Ref.~\onlinecite{thio88}, and, as far as we understand, does not follow from neutron scattering experiments. 

\subsection{Gyromagnetic ratio}
The assumption that the copper spins in LCO are accurately described by the usual electron gyromagnetic ratio $g=2$ is not quite correct. Let us therefore consider the renormalization of $g$ due to the spin-orbit interaction. The electronic configuration of the Cu in La$_2$CuO$_4$ is $3d^9$, with one active hole per Copper atom. In the case of non-zero orbital angular momentum ${\vec l}$, the spin-orbit coupling and the interaction with the magnetic field read
\begin{equation*}
H_{\text{int}}=\lambda {\vec l} \cdot {\vec S} - \mu_{B} \left({\vec l}+ 2 {\vec S}\right) {\vec B} \ ,
\end{equation*}
with $\lambda \approx -0.09 \ eV$, see e.g., Ref.~\onlinecite{pennington89}. Considering the  terms involving the orbital angular momentum as perturbations, we derive an effective Hamiltonian of the form
\begin{equation*}
H_{\text{eff}}=-g_{\alpha} \ \mu_{B} \  S^\alpha B^\alpha \ ,
\end{equation*}
with a generalized gyromagnetic tensor $g$ which is diagonal in the axes $a$, $b$, and $c$. The ground state and the first three excited states of La$_2$CuO$_4$ are summarized in Tab.~\ref{tab:energylevels}. For simplicity, we neglect the small orthorhombic distortion here.
\begin{table}
\centering \caption{\label{tab:energylevels}
\emph{Lowest-lying energy levels and wave functions of La$_2$CuO$_4$. For simplicity, we neglect the small orthorhombic distortion. $Y_{l,m}$ are spherical harmonics.}}
\begin{ruledtabular} \begin{tabular}{ccc}
Orbital & Wavefunction & Energy (eV) \\ \hline \\
$ 3d_{x^2-y^2}$ & $ \left| Y_{2,2}+Y_{2,-2} \right\rangle$ & $0$ \\[0.2cm]
$ 3d_{xy} $ & $  \left| Y_{2,2}-Y_{2,-2} \right\rangle$ & 1.35\footnotemark[1] \\[0.2cm]
$ 3d_{3z^2-r^2} $ & $  \left| Y_{2,0} \right\rangle$ & 1.5\footnotemark[2] \\[0.2cm]
$ 3d_{xz} $ & $  \left| Y_{2,1}+Y_{2,-1} \right\rangle$ & 1.70\footnotemark[2] \\
$ 3d_{yz} $ & $  \left| Y_{2,1}-Y_{2,-1} \right\rangle$ & 1.70\footnotemark[2]
\end{tabular} \end{ruledtabular}
\footnotetext[1]{Refs.~\onlinecite{liu93,salamon95}.}
\footnotetext[2]{Ref.~\onlinecite{kuiper98}.}
\end{table}
Second order perturbation theory yields corrections to the energy given by
\begin{equation*}
\delta E_{\sigma,\sigma'} = -2 \lambda \mu_{B} \sum_{\left| n \right\rangle \ne 0, \tilde \sigma} \frac{ \left\langle 0,\sigma \right|  {\vec l}  \cdot {\vec B} \left| n, \sigma\right\rangle
\left\langle n,\tilde \sigma \right| {\vec l}\cdot {\vec S} \left| 0,\sigma' \right\rangle}{E_{0}-E_{n}} \ ,
\end{equation*}
where the factor $2$ gives the correct multiplicity of the considered processes. With the knowledge of the non-zero matrix elements of the orbital angular momentum operator
\begin{align*} 
\left\langle 3d_{xy} \right| l^z \left| 3d_{x^2-y^2} \right\rangle &= 2 \ , \\
\left\langle 3d_{yz} \right| l^x \left| 3d_{x^2-y^2} \right\rangle &= 1 \ , \\
\left\langle 3d_{xz} \right| l^y \left| 3d_{x^2-y^2} \right\rangle &= i \ ,
\end{align*}
it is straightforward to calculate the gyromagnetic ratio
\begin{align} \label{eq:gyro}
g_{\perp} &=2-\frac{2 \lambda}{1.7} \approx 2.1 \ , \nonumber \\ 
g_{\parallel} &=2-\frac{8 \lambda}{1.35} \approx 2.5 \ .
\end{align}
We use these values for further calculations.

\section{Field dependence of the spin-wave spectrum\label{sec:spectrum}}
Depending on its direction, an applied magnetic field can lead to a reorientation of the staggered magnetization. Similar to the approach followed in Sec.~\ref{sec:coupling} in the absence of a magnetic field, we proceed as follows. We first find the static ground state orientation of the order parameter and then consider excitations, described as small deviations from the ground state configuration. This allows us to determine the spin-wave spectrum in the presence of uniform magnetic fields. The excitations, denoted $n_{1}$ and $n_{2}$ are perpendicular to the static ground state orientation and correspond to the in- and out-of-plane modes at zero magnetic field.

In order to keep the formalism as light as possible, we chose to restrict the staggered magnetization to be uniform within a given copper-oxide layer, i.e., ${\vec n}_{l}\left({\bf r}\right) = {\vec n}_{l}$. This approach has two advantages: Firstly, we do not have to carry the in-plane momenta through the derivations, which greatly improves the readability of the presentation and secondly, it is straightforward to remove this constraint, by restoring the momenta in the final answer, by replacing 
$\omega_{j}^2 \rightarrow \omega_{j}^2 - c^2 k^2$. 

\subsection{Magnetic field along $a$\label{sec:fielda}}
For a magnetic field applied along the $a$ axis, i.e., ${\vec B}=B {\vec e}_{a}$, the Lagrangian~(\ref{eq:lagrangian}) reads
\begin{multline*} 
{\cal L} = \sum_{l=1}^L \left\{ \dot{\vec n}_{l}^2 - D^2 \left(n_{l}^{a}\right)^2- \Gamma_{c} \left(n_{l}^c\right)^2 \right. \\ \left. + B^2\left[\left(n_{l}^{b}\right)^2+\left(n_{l}^c\right)^2\right] - \gamma^2 {\vec n}_{l} \cdot {\vec n}_{l+1} \right\} \ .
\end{multline*}
Clearly, a field along the $a$ direction does not change the orientation of the staggered magnetization and the ground state energy is equal to
\begin{equation*}
E^{a}=-\frac{\chi_{\perp}}{2} \left( B^2+\gamma^2\right) \ .
\end{equation*}
Considering small deviations from this state, by introducing 
\begin{equation} \label{eq:na}
{\vec n}_{l} = \left(n_{l,1}, \left(-1\right)^l \sqrt{1-n_{l,1}^2-n_{l,2}^2},n_{l,2}\right) \ ,
\end{equation}
together with the corresponding time derivative
\begin{equation*}
\dot{\vec n}_{l} \approx \left( \dot{n}_{l,1}, 0, \dot{n}_{l,2}\right) \ ,
\end{equation*}
we find~\cite{comment2}
\begin{multline*}
{\cal L}= \sum_{l=1}^L \left\{ \dot{n}_{l,1}^2 + \dot{n}_{l,2}^2 - n_{l,1}^2 \left(D^2 +B^2+\gamma^2\right) \right. \\ \left. - n_{l,2}^2 \left(\Gamma_{c}+\gamma^2\right)  - \gamma^2\left(n_{l,1} n_{l+1,1}+n_{l,2} n_{l+1,2}\right) \ \right\} \ ,
\end{multline*}
which after Fourier transformation~(\ref{eq:fourier}) becomes
\begin{multline*}
{\cal L}= \sum_{q} \left\{ \dot{n}_{q,1}^* \dot{n}_{q,1} + \dot{n}_{q,2}^* \dot{n}_{q,2} \right. \\Ê\left. - n_{q,1}^* n_{q,1} \left[D^2 +B^2+\gamma^2\left(1+\cos q\right)\right] \right. \\ \left. - n_{q,2}^*n_{q,2} \left[\Gamma_{c}+\gamma^2\left(1+\cos q\right)\right]\right\} \ .
\end{multline*}
With the time dependence~(\ref{eq:timedep}), the equations of motion are decoupled and we easily deduce the spin-wave spectrum (restoring the in-plane momentum as explained at the beginning of Sec.~\ref{sec:spectrum})
\begin{align} \label{eq:spectruma}
\omega_{1}^2 &= c^2 k^2 + D^2 +B^2+\gamma^2\left(1+\cos q\right) \ , \nonumber \\
\omega_{2}^2 &= c^2 k^2 + \Gamma_{c}+\gamma^2\left(1+\cos q\right) \ .
\end{align}
\begin{figure}
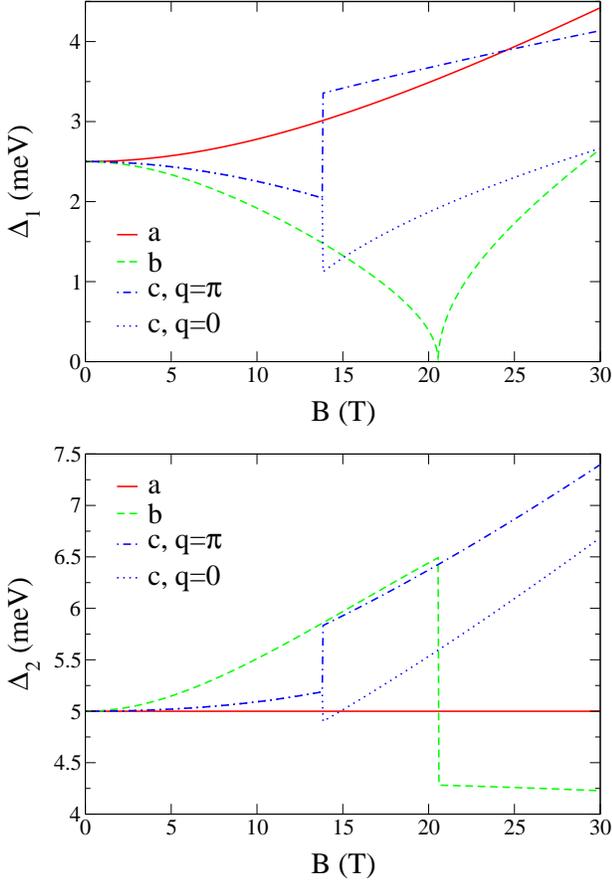

\includegraphics[width=0.45\textwidth,clip]{fig1a.eps} \\[0.2cm]
\includegraphics[width=0.45\textwidth,clip]{fig1b.eps}
\caption{\emph{(Color online). Field-dependence of the spin-wave gaps for different directions of the uniform magnetic field, labeled by the corresponding axis. $\Delta_{1}$ can be identified as the in-plane gap, while $\Delta_{2}$ corresponds to the out-of-plane gap at zero field. These plots are based on the parameters derived in Sec.~\ref{sec:model}, i.e., $D=2.5\ meV$, $\sqrt{\Gamma_{c}}=5 \ meV$, $\gamma^2=5 \left(meV\right)^2$, $g_{\parallel}=2.1$, and $g_{\perp}=2.5$. For a field $B>B_{\text{flop}}^c$ applied along the c axis, we show the curves for momenta $q=0$ and $q=\pi$. A comparison with recent experiments is presented in Fig.~\ref{fig:gaps2}.} \label{fig:gaps}}
\end{figure}
In the absence of a magnetic field, we find the usual in- and out-of-plane gaps at momentum $\left({\bf k},q\right)=\left(0,\pi\right)$, see Eq.~(\ref{eq:spectrum0}). For an increasing field applied along the $a$ axis, the in-plane gap hardens, while the out-of-plane gap remains constant. The field-dependence of the spin-wave gaps is presented in Fig.~\ref{fig:gaps}.

\subsection{Magnetic field along $b$ \label{sec:fieldb}}
\begin{figure}
\includegraphics[width=0.4\textwidth,clip]{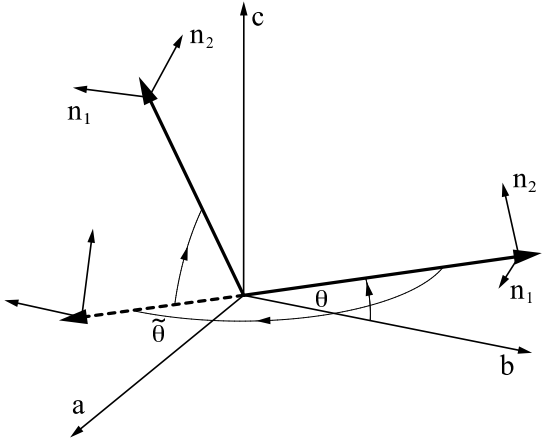}
\caption{\emph{A magnetic field along the $b$ direction leads to a rotation of the staggered magnetization first around the $a$ axis and then around the $b$ axis.} \label{fig:lattice}}
\end{figure}

For a magnetic field applied along the $b$ axis, i.e., ${\vec B}=B {\vec e}_{b}$, the Lagrangian~(\ref{eq:lagrangian}) reads
\begin{multline} \label{eq:lagrangianb}
{\cal L} = \sum_{l=1}^L \left\{ \dot{\vec n}_{l}^2 -2 \dot{\vec n}_{l} \left( {\vec n}_{l} \times {\vec B}\right)-D^2 \left(n_{l}^{a}\right)^2- \Gamma_{c} \left(n_{l}^c\right)^2 \right. \\Ê\left. +B^2\left[\left(n_{l}^{a}\right)^2+\left(n_{l}^c\right)^2\right] +2 B D n_{l}^c -  \gamma^2 {\vec n}_{l} \cdot {\vec n}_{l+1} \right\} \ .
\end{multline}
In this case, the magnetic field first leads to a rotation of the ${\vec n}$-field around the $a$ axis, as shown in Fig.~\ref{fig:lattice}. The angle $\theta$ is given by (see also Ref.~\onlinecite{chovan00})
\begin{equation} \label{eq:theta1}
\sin \theta = \frac{BD}{\Gamma_{c}+2 \gamma^2-B^2} \ ,
\end{equation}
and the ground state energy is equal to
\begin{equation*}
E_{1}^b=-\frac{\chi_{\perp}}{2} \left(\gamma^2+\frac{\left(B D\right)^2}{\Gamma_{c}+2 \gamma^2-B^2}\right) \ .
\end{equation*}
At $B_\text{crit}^b=D\approx 21 T$, it becomes energetically favorable to have the largest component of the staggered magnetization directed along the $a$ axis, see Eq.~(\ref{eq:lagrangianb}). For even larger fields, the order parameter is then further rotated towards the $c$ axis, with an angle $\tilde \theta$ between ${\vec n}$ and the $a$ axis equal to (see also Ref.~\onlinecite{chovan00})
\begin{equation} \label{eq:thetatilda}
\sin \tilde \theta = \frac{BD}{\Gamma_{c}-D^2+2 \gamma^2} \ .
\end{equation}
In this case, the ground state energy is given by
\begin{equation*}
E_{2}^b=-\frac{\chi_{\perp}}{2} \left( \gamma^2+B^2-D^2+\frac{\left(B D\right)^2}{\Gamma_{c}+2 \gamma^2-D^2} \right) \ .
\end{equation*}
Since ${\vec n}^2=1$, we find that for a field of around $B\approx90 T$, the staggered magnetization is completely aligned along the $c$ direction.

\subsubsection{$B < B_\text{crit}^b$}
Let us now turn to the field-dependence of the spin-wave spectrum. In the rotated coordinate system, we can express the components of ${\vec n}$ on even and odd sites as
\begin{equation*}
{\vec n_{l}} = 
\begin{pmatrix} 
n_{l,1} \\[0.2cm]
\left(-1\right)^l \left( \sqrt{1-n_{l,1}^2-n_{l,2}^2} \cos \theta - n_{l,2} \sin \theta\right) \\[0.2cm]
\sqrt{1-n_{l,1}^2-n_{l,2}^2} \sin \theta + n_{l,2} \cos \theta 
\end{pmatrix} \ .
\end{equation*}
Its time-derivative is thus given by
\begin{equation*}
\dot{\vec n}_{l} \approx
\begin{pmatrix} 
\dot{n_{l,1}} \\[0.2cm]
\left(-1\right)^{l+1} \dot{n}_{l,2} \sin \theta \\[0.2cm]
\dot{n}_{l,2} \cos \theta 
\end{pmatrix} \ .
\end{equation*}

Using the above decomposition of the staggered magnetization, together with the expression for the angle of rotation~(\ref{eq:theta1}), we find the Lagrangian~\cite{comment2}
\begin{multline*}
{\cal L}= \sum_{q} 
\left\{ \dot{n}_{q,1}^* \dot{n}_{q,1} + \dot{n}_{q,2}^* \dot{n}_{q,2} \right. \\Ê\left. - n_{q,1}^*n_{q,1} \left[D^2 -B^2+\gamma^2\left(1+\cos q\right)\right] \right. \\ \left.
- n_{q,2}^*n_{q,2} \left[\left(\Gamma_{c}-B^2+\gamma^2\left(1+\cos q\right)\right) \cos 2 \theta + B D \sin \theta \right] \right. \\Ê\left. + 2 B \cos \theta  \left(\dot{n}_{q,1}^* \dot{n}_{q,2} -\dot{n}_{q,2}^* \dot{n}_{q,1}\right) \right\} \ .
\end{multline*}
Note that in contrast to the previous case (Sec.~\ref{sec:fielda}), the two modes are now coupled through the last term in the above equation. Substituting the time dependence~(\ref{eq:timedep}), the Euler-Lagrange equations read
\begin{equation*}
\begin{pmatrix} 
-\omega^2+\xi_{1}^2 & 2 i \omega B \cos \theta \\[0.3cm]
-2 i \omega B \cos \theta &Ê-\omega^2+\xi_{2}^2 
\end{pmatrix}
\begin{pmatrix}
n_{q,1}^0 \\[0.3cm]
n_{q,2}^0
\end{pmatrix}
= 0 \ ,
\end{equation*}
where
\begin{align*}
\xi_{1}^2 &= D^2 -B^2+\gamma^2\left(1+\cos q\right) \ , \\
\xi_{2}^2 &= \left[\Gamma_{c}-B^2+\gamma^2\left(1+\cos q\right)\right] \cos 2 \theta + B D \sin \theta \ .
\end{align*}
From here, one finds the spin-wave spectrum for magnetic fields $B<D$
\begin{multline} \label{eq:spectrumb1}
\omega_{j}^2 =c^2 k^2 +  \frac{1}{2}\left(\xi_{1}^2+\xi_{2}^2+4 B^2 \cos^2 \theta  \right. \\ \left. \pm 
\sqrt{ \left(\xi_{1}^2 +\xi_{2}^2 + 4 B^2 \cos^2 \theta \right)^2-4  \xi_{1}^2 \xi_{2}^2} \right) \ .
\end{multline}
The minimum of the spin-wave gaps is found at $\left({\bf k},q\right)=\left(0,\pi\right)$ and their dependence on the magnetic field is plotted in Fig.~\ref{fig:gaps}.

\subsubsection{$B>B_\text{crit}^b$}
For magnetic fields $B>D$, the staggered magnetization is contained in $ac$ plane. Following the same approach as below the critical field, we expand the order parameter as
\begin{equation*}
{\vec n_{l}} = 
\begin{pmatrix}
\left(-1\right)^l \left( \sqrt{1-n_{l,1}^2-n_{l,2}^2} \cos \tilde \theta - n_{l,2} \sin \tilde \theta\right) \\[0.2cm]
-n_{l,1} \\[0.2cm]
\sqrt{1-n_{l,1}^2-n_{l,2}^2} \sin \tilde \theta + n_{l,2} \cos \tilde \theta 
\end{pmatrix} \ ,
\end{equation*}
its time derivative being equal to
\begin{equation*}
\dot{\vec n}_{l} = 
\begin{pmatrix}
- \dot{n}_{l,2} \sin \tilde \theta \\[0.2cm]
-\dot{n}_{l,1} \\[0.2cm]
\dot{n}_{l,2} \cos \tilde \theta 
\end{pmatrix} \ .
\end{equation*}
After Fourier transformation~(\ref{eq:fourier}) and substitution of the angle $\tilde \theta$~(\ref{eq:thetatilda}), the Lagrangian~(\ref{eq:lagrangianb}) reads~\cite{comment2}
\begin{multline*}
{\cal L}= \sum_{q} 
\left\{ \dot{n}_{q,1}^* \dot{n}_{q,1} + \dot{n}_{q,2}^* \dot{n}_{q,2} \right. \\ \left.- n_{q,1}^*n_{q,1} \left[B^2-D^2+\gamma^2\left(1+\cos q\right)\right] \right. \\ \left.
- n_{q,2}^*n_{q,2} \left[\left(\Gamma_{c}-D^2 +\gamma^2\left(1+\cos q\right)\right) \cos 2 \tilde \theta + B D \sin \tilde \theta \right] \right\} \ .
\end{multline*}
For fields $B>D$, the spin-wave dispersions are therefore given by
\begin{align} \label{eq:spectrumb2}
\omega_{1}^2 &= c^2 k^2 + B^2-D^2+\gamma^2\left(1+\cos q\right) \ , \nonumber \\Ê
\omega_{2}^2 &= c^2 k^2 + \left[\Gamma_{c}-D^2 +\gamma^2\left(1+\cos q\right)\right] \cos 2 \tilde \theta \nonumber \\ &\hspace*{1cm}+ B D \sin \tilde \theta \ .
\end{align}
In the presence of moderate magnetic fields, such that $\cos 2 \tilde \theta >0$, which roughly corresponds to $B \lesssim 65\ T$, the minima of the spin-wave gaps are found at $\left({\bf k},q\right)=\left(0,\pi\right)$. As soon as $\tilde \theta > \pi/4$, i.e., for extremely high fields, a change in momentum of the out-of-plane mode from $\pi$ to $0$ is favorable.

As illustrated in Fig.~\ref{fig:gaps}, the in-plane gap softens for $B<D$ and eventually vanishes exactly at the critical field. By further increasing the magnetic field, the gap hardens again, quickly becoming larger than the zero-field value $D$. In contrast, the out-of-plane gap first hardens and then jumps at the transition to decrease very slowly for fields $B>D$. It eventually vanishes in the fully polarized state.

\subsection{Magnetic field along $c$ \label{sec:fieldc}}
Lastly, we study the effects of a uniform magnetic field parallel to the $c$ axis, i.e., ${\vec B}=B {\vec e}_{c}$. The Lagrangian~(\ref{eq:lagrangian}) then takes the form~\cite{comment2}
\begin{multline} \label{eq:lagrangianc}
{\cal L} = \sum_{l=1}^L \left\{ \dot{\vec n}_{l}^2 - D^2 \left(n_{l}^{a}\right)^2- \Gamma_{c} \left(n_{l}^c\right)^2 \right. \\Ê\left. +B^2\left[\left(n_{l}^{a}\right)^2+\left(n_{l}^b\right)^2\right] -
 2 B D n_{l}^b -  \gamma^2 {\vec n}_{l} \cdot {\vec n}_{l+1} \right\} \ .
\end{multline}
We would like to point out that this case has been analyzed earlier in Ref.~\onlinecite{chovan00}. Nevertheless, we reconsider it here to facilitate comparison with experiments in next section. For a magnetic field applied along the $c$ axis, there is competition between the interlayer coupling term, which leads to an antiparallel alignment of the staggered magnetization (still oriented along the $b$ axis) and the term linear in the magnetic field, for which a parallel orientation is energetically favorable. The mechanism of the experimentally observed spin-flop transition follows therefore naturally from the above Lagrangian.

Let us first calculate the critical field at which the spin-flop transition takes place. By comparing the energy of the parallel
\begin{equation*}
E_{1}^c =-\frac{\chi_{\perp}}{2} \left(\gamma^2+B^2\right)
\end{equation*}
with the antiparallel configuration 
\begin{equation*}
E_{2}^c =-\frac{\chi_{\perp}}{2} \left(-\gamma^2+B^2+2 B D \right) \ , 
\end{equation*}
we find
\begin{equation*}
B_\text{flop}^c = \frac{\gamma^2}{D} \approx 14\ T \ ,
\end{equation*}
using the value of $\gamma^2$ given by Eq.~(\ref{eq:gamma}). This estimation is pretty close to the most recent experimental determination of the critical field of around $11.5\ T$, see Ref.~\onlinecite{reehuis06}, but substantially larger than the measurement by Gozar {\it et al.}~\cite{gozar04} ($6.5\ T$). 

Below the spin-flop transition (for antiparallel alignment), we use the expansion of the ${\vec n}$-field given in Eq.~(\ref{eq:na}). In momentum representation, the Lagrangian~(\ref{eq:lagrangianc}) then reads~\cite{comment2}
\begin{multline*}
{\cal L} = \sum_{q} \left\{ \dot{n}_{q,1}^* \dot{n}_{q,1} + \dot{n}_{q,2}^* \dot{n}_{q,2}  \right. \\ \left. - n_{q,1}^*n_{q,1}  \left[D^2+ \gamma^2 \pm \sqrt{B^2 D^2+\gamma^4 \cos^2 q}\right]  \right. \\Ê
\left. - n_{q,2}^* n_{q,2} \left[ \Gamma_{c} + B^2+\gamma^2 \pm \sqrt{B^2 D^2+\gamma^4 \cos^2 q} \right] \right\} \ ,
\end{multline*}
with two branches, because the momenta are limited to the reduced Brillouin zone, i.e., $q\in \left[-\pi/2,\pi/2\right)$. After restoration of the in-plane momentum, we find the spin-wave spectrum for fields $B<B_{\text{flop}}^c$
\begin{align} \label{eq:spectrumc1}
\omega_{1}^2 &= c^2 k^2 +D^2+ \gamma^2 \pm \sqrt{B^2 D^2+\gamma^4 \cos^2 q} \ , \nonumber \\
\omega_{2}^2 &= c^2 k^2+ \Gamma_{c} + B^2+\gamma^2 \pm \sqrt{B^2 D^2+\gamma^4 \cos^2 q} \ .
\end{align}
The minima of the gaps are therefore located in the lower branch at momentum $\left({\bf k},q\right)=\left(0,0\right)$. These gaps have been found earlier in Ref.~\onlinecite{chovan00}.

Above the spin-flop transition (for parallel alignment of the order parameter), we omit the oscillating factor in the expansion of the staggered magnetization~(\ref{eq:na}) and find the Lagrangian~\cite{comment2}
\begin{multline*}
{\cal L} = \sum_{q} \left\{ \dot{n}_{q,1}^* \dot{n}_{q,1} + \dot{n}_{q,2}^* \dot{n}_{q,2} \right. \\Ê\left. - n_{q,1}^*n_{q,1} \left[D^2+BD-\gamma^2\left(1-\cos q\right)\right] \right. \\ \left.
- n_{q,2}^*n_{q,2} \left[\Gamma_{c}+B^2 +BD -\gamma^2\left(1-\cos q\right) \right] \right\} \ .
\end{multline*}
The excitation spectrum is thus equal to (see also Ref.~\onlinecite{chovan00})
\begin{align} \label{eq:spectrumc2}
\omega_{1}^2 &= c^2 k^2 + D^2+BD-\gamma^2\left(1-\cos q\right) \ , \nonumber \\
\omega_{2}^2 &= c^2 k^2 + \Gamma_{c}+B^2 +BD -\gamma^2\left(1-\cos q\right) \ .
\end{align}
Strictly speaking, the minimum of the gaps is found at $\left({\bf k},q\right)=\left(0,\pi\right)$. This implies that there is a change in momentum of the low-lying excitations at the spin-flop transition. However, the comparison with experiments presented in Fig.~\ref{fig:gaps2} suggests that Raman scattering actually follows the excitation with momentum $q=0$.

\section{Comparison with experiments\label{sec:gaps}}
Let us now compare our findings with recent measurements by Gozar and co-workers~\cite{gozar04}, who reported the field dependence of the spin-wave gaps in LCO observed in Raman spectroscopy. As explained in detail in Ref.~\onlinecite{silvaneto05}, the in-plane mode can be observed for all three directions of the field, whereas the out-of-plane gap is only detectable when the magnetic field is applied along the orthorhombic $b$ direction.
\begin{figure}
\includegraphics[width=0.45\textwidth,clip]{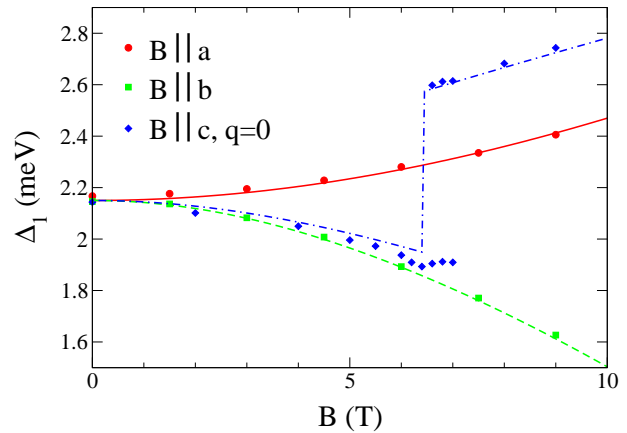}
\caption{\emph{(Color online). Field-dependence of the in-plane spin-wave gaps for different directions of the applied magnetic field. Symbols represent the experimental observations shown in Fig.~2 of Ref.~\onlinecite{gozar04}, while the lines correspond to our theoretical calculations based on the parameters~(\ref{eq:parameters}) and the gyromagnetic tensor~(\ref{eq:gyro}). Note that these values are slightly different from those used in Fig.~\ref{fig:gaps}.} \label{fig:gaps2}}
\end{figure}
At this stage, it seems appropriate to comment on the uncertainties related to the basic parameters that enter the expressions for the spin-wave gaps, namely the anisotropies (gaps at zero field) and the interlayer coupling. The plots presented in Fig.~\ref{fig:gaps} are based on the generally accepted values for these parameters. However, their precise values vary from one experiment to another, presumably due to slightly different samples. From the experiment under consideration~\cite{gozar04}, we deduce 
\begin{align} \label{eq:parameters}
D &= 2.15 \ meV \nonumber \\ 
\gamma^2 &= B_\text{flop}^c D = 2 \left(meV\right)^2 \ .
\end{align}
While the value of the in-plane gap is in very good agreement with Ref.~\onlinecite{keimer93}, the huge difference in $\gamma^2$ is quite puzzling. Nevertheless, in Fig.~\ref{fig:gaps2}, we have reproduced the measurements of the in-plane gaps presented in Fig.~2 of Ref.~\onlinecite{gozar04} together with the results of our theoretical calculations based on the parameters given above and our estimates of the gyromagnetic ratios~(\ref{eq:gyro}). The agreement between theory and experiments is quite remarkable. From the out-of plane mode observed in Fig.~4 of Ref.~\onlinecite{gozar04}, we conclude that a field applied along the $b$ axis leads to a slight hardening of the gap. This is also in good agreement with our result shown in Fig.~\ref{fig:gaps}.

\section{Conclusion\label{sec:conclusion}}
To conclude, we analyze the evolution of the spin-wave spectrum in La$_2$CuO$_4$ in a uniform magnetic field. We find that a field applied along the orthorhombic $b$ direction leads to a rotation of the staggered magnetization first in the $bc$ and then in the $ac$ plane, until the order parameter is completely aligned along the $c$ direction. In this case, the field-dependence of the spin-wave spectrum is given by Eqs.~(\ref{eq:spectrumb1}) and (\ref{eq:spectrumb2}). In contrast, for a perpendicular field, there is an ordinary spin-flop transition induced by the competition between the magnetic field and the interlayer coupling, which leads to the spin-wave modes given by Eqs.~(\ref{eq:spectrumc1}) and (\ref{eq:spectrumc2}). In the case where the magnetic field is directed along the $a$ axis, the staggered magnetization remains aligned along the $b$ direction and the evolution of the spin-wave dispersion is given by Eq.~(\ref{eq:spectruma}). A comparison with recent measurements of the field-dependence of the spin-wave gaps shows a beautiful agreement between theory and experiments. We hope that this work can stimulate new neutron scattering experiments that could reconcile the very different spin-flop transition fields measured in experiments~\cite{gozar04,reehuis06}.

After completion of this work, we discovered two very recent preprints by Benfatto {\it et al.}~\cite{benfatto06a,benfatto06b} which among other questions address the same problem as we do in this paper. We would especially like to draw the attention to Ref.~\onlinecite{benfatto06b}, which contains new measurements of the out-of-plane gap and an insightful description of the Raman response. Comparing our results, we find that they are almost identical, except in the case where a strong magnetic field is applied along the $b$ direction, see Eq.~(\ref{eq:spectrumb2}). While Benfatto and Silva Neto~\cite{benfatto06a} predict a constant out-of plane gap for $B>D$, our results shown in Fig.~\ref{fig:gaps} indicate a slow softening of the gap, which eventually vanishes at the point where the order parameter is completely aligned along the $c$ axis.

\acknowledgments{We are grateful to A. Lavrov, Y. Ando, A. Gozar, and G. Khaliullin for valuable discussions.}


\begin{thebibliography}{24}
\bibitem{comment1} There are two ways of imposing ${\vec n}_{l}^2=1$: We always use the ``hard'' constraint, by considering two small transverse modes $n_{l,j}$, which determine the longitudinal component of ${\vec n}_{l}$ as $\sqrt{1-n_{l,1}^2-l_{l,2}^2}$. An alternative (``soft'') way consists of introducing a very large but finite longitudinal mass. However, this is just a technical difference that cannot influence the physical conclusions in the low-energy sector.
\bibitem{comment2} We neglect terms independent of $n_{l,j}$ and gauge out contributions linear in $\dot{n}_{l,j}$.
\end{thebibliography}
\end{document}